\documentclass[onecolumn,epsfig,showpacs,preprintnumbers,amsmath,amssymb,superscriptaddress]{revtex4}
\usepackage{graphicx}
\usepackage{epsfig}
\usepackage{dcolumn}
\usepackage{bm}
\usepackage[all]{xy}
\usepackage{ifthen}
\usepackage{subfigure}

\begin{document}

\title{Travel and Tourism: into a Complex Network}

\author{J. I. L. Migu\'ens}
\email{joana.miguens@ua.pt}
\affiliation{Economics, Management and Industrial Engineering Department, Aveiro University, 3810 - 193 Aveiro, Portugal}%

\author{J. F. F. Mendes}
\email{jfmendes@ua.pt} \affiliation{ Physics Department, Aveiro
University, 3810 - 193 Aveiro, Portugal}

\date{\today}


\begin{abstract}

It is discussed how the worldwide tourist arrivals, about 10\% of
world's domestic product, form a largely heterogeneous and directed
complex network. Remarkably the random network of connectivity is
converted into a scale-free network of intensities. The importance
of weights on network connections is brought into discussion. It is
also shown how strategic positioning particularly benefit from
market diversity and that interactions among countries prevail on a
technological and economic pattern, questioning the backbones of
traveling driving forces.

\end{abstract}

\maketitle

\section{\label{sec:level1}Introduction}

The movement of tourists on a worldwide scale is responsible for a
traveling mobility of hundred millions tourist arrivals every year,
representing the largest movement of humans ever out of their usual
environment, strongly influencing local, regional, national and
international economies, being one of the fastest growing economic
sector. Tourism is a consequence and a dynamic force on the
integration of world trade and markets, forming the global economy.
But how is this integration evolving? However, regardless the
crucial role of tourism, there is a lack of quantitative
considerations of its flows, although it is essential for
understanding the self-organization of human traveling patterns,
and global wealth net flows.

Research on social networks has around $50$ years, empirical and
theoretically, partly because social life is relational
\cite{milgram,granovetter}. These studies contributed much for the
clarity of the importance of relational systems. Such networks are
represented as a set of \emph{nodes} denoting people, companies, or
other social actors, which are joined by \emph{edges} the patterns
of the relational structure, representing friendships, partnerships,
collaborations, etc. The increasing availability of real networks
data and the nowadays capacity of analyzes large data, have enhanced
new analytic methods to characterize networks, extending our
knowledge on the description of these systems
\cite{albert-2002-74,goltsev-2002-51,newman-2003-45}.

A large variety of real world systems are structured in the form of
networks, from social, biological , economic, infrastructure and
information networks
\cite{mendesbook,Strogatz:ExploringComplexNetworks,wassermanSNA,albert-2002-74,jeong-2001-411}.
Network theory have been build up largely from observation of the
properties of many real world networks, and by comparatione of their
structures. Due to the interesting results,  research on complex
networks has significantly increased during the last years. These
methods have been applied to a wide variety of real world networks,
like airline connections \cite{air}, financial relations
\cite{marketInvestment,finantial,tibely-2006-370}, companies
partnerships, ecological networks, movies actors, world trade, WWW
\cite{barabasiInternet}, scientific collaboration network
\cite{barabasi-2002-311}, human acquaintance patterns
\cite{newman2001b}, among others
\cite{travel,Montis2005,barrat-2004-101}.

Tourism is one of the fastest growing economic sectors, representing
more than $10.2\%$ of world GDP, and reached a record of
international tourist arrivals with $763$ million in $2004$.
Different theoretical perspectives on tourism recognize clusters and
networks as one of the main competitive factors in tourism.
An organizational perspective applying social network analysis
was introduced in $1996$ \cite{costa}.
Since then tourism researchers have been introducing network
analysis on measuring international business, drive tourism
\cite{tourismNetworks,tourismNetworksDriveTourism}, and
more recently on electronic tourism with a
web graph structure of a destination \cite{rodolfo,rodolfo2}.

The research aims to show that network analysis has appropriate
methods to study the worldwide tourist arrivals, as a network
system. This paper is organized as follows. An introduction to the
international tourism on section \ref{sec:introITN} and to network
theory on section \ref{sec:NT}. The empirical analysis (section
\ref{sec:empiricalAnalysis}) focuses on topological and weighted
analyze (section \ref{sec:topologyWeights}) and degree correlation
(section \ref{sec:degreeCorrelations}). Conclusion are drawn on
section \ref{sec:conclusion}.

\subsection{Worldwide Tourist Arrivals}\label{sec:introITN}
\begin{figure}[!hbp]
  \includegraphics[width=10cm]{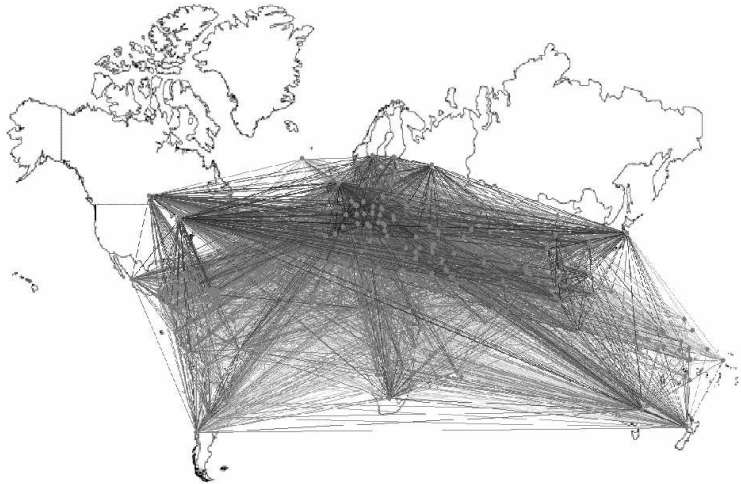}\\
  \caption{{\small Worldwide Tourism is a complex network
  The map of the world tourism network is displayed with an
  exponential grey scale according to the intensity of connections.
  }
  }\label{fig:worldMap}
\end{figure}

In this research we use a network approach to study international
tourism on the year of $2004$. International tourist arrivals
reached a record of $763$ million in $2004$ (see Fig.
\ref{fig:worldMap}). The international arrival of tourist is yearly
measured by the World Tourism Organization (WTO, the major
intergovernmental body concerned with tourism) over $208$ countries
and territories around the world \cite{yearbook}. Worldwide earnings
on international tourism reached in $2004$ a new record value of US
$623$ billion.

International tourist arrivals are analyzed to study \emph{inbound}
tourism and \emph{outbound }tourism. \emph{Inbound} tourism,
involving the non-residents received by a destination country from
the point of view of that destination. \emph{Outbound} tourism,
involving residents traveling to another country from the point of
view of the country of origin.

This study provides
information on the role played by network theory on the structure of
international travel flows.

\subsection{Network Theory}\label{sec:NT}
We argue that network theory provides an explanatory framework of
interrelationships of how countries as tourism destinations
interact, relate, and evolve. Techniques and indicators of network
theory are introduced for measuring relation in the international
network theory.

Most complex networks share common properties that have common
underlying structural principles
\cite{mendesbook,barabasiInternet,goltsev-2002-51}. Real world
networks have been shown to display structural properties different
from random graphs model.

The centrality of nodes on a network is of primary importance, as
more competitive nodes have better strategic positions
\cite{burt,wassermanSNA}. Several measures of centrality have been
developed, like degree centrality, closeness, betweenness,
eigenvector centrality, information centrality, among others.
Centralization refers to the extent to which a network revolves
around a single node, and also to the propensity of the node to
diffuse information, knowledge or infections.

Degree centrality is one of the most used measures of node
prominence \cite{freeman}. The degree of a node, represents the
prominence of a node, and equals the number of edges connected to
it. The statistical characterization of real networks displays a
large number of node degrees, $k$, and the appearance of hubs, nodes
with large degree \cite{goltsev-2003-67}. Additionally these
networks show a scale-free degree distribution , characterized by a
power-law behavior $P(k) \sim k^{-\theta}$
\cite{barabasiInternet,mendesbook}.

Despite the wide range of application, complex networks have
developed to the characterization different topological networks,
undirected and directed, unweighted and weighted
\cite{barrat-2004-101}. The techniques firstly applied to undirected
and unweighted networks are lately adapted to weighted and/or
directed networks \cite{newman-2003-45,barrat-2004-92,park-2006-74}.
Topological properties have a very strong influence on propagation
of knowledge and disease, as well as on robustness and vulnerability
\cite{barabasi1999,mendesbook}. Despite the importance of
topological issues, weighted analyzes characterize the heterogeneity
of weights and non-trivial correlation
\cite{barrat-2004-101,Montis2005}.

Directed edges are considered when the edge from node $i$ to node
$j$ ($i \to j$) is different of the edge from node $j$ to node $i$
($j \to i$). Many real networks are also weighted networks, in the
case of social networks it is often relevant to assign a weight
(strength) to each edge, measuring how good or strong is a
relationship \cite{granovetter,newman2001b}.

\section{Empirical Analysis}\label{sec:empiricalAnalysis}
Follows the empirical study of worldwide tourist flows,
for the year of $2004$. Topological and
weighed analyzes are performed on section \ref{sec:topologyWeights}.
The way tourism destinations
couple together is analyzed through degree-degree correlations on section
\ref{sec:degreeCorrelations}.

\subsection{Topology and Weights}\label{sec:topologyWeights}
We used the data gathered by WTO over these $208$ where countries
and territories are considered \emph{nodes}, N, and an \emph{edge}
exists from node $i$ to node $j$ when there are tourists from
country $i$ to country $j$. Notice that the network is directed, the
edge from $i$ to $j$ is different from the edge from $j$ to $i$,
respectively $i \to j$ and $j \to i$. On our case we have $5775$
edges, L, -- representing arrivals of tourists from one country to
another, on the year of $2004$. For an unweighted network node is
one of the most studied centrality measures in social network
analysis.

An important statistical property to directed networks is
reciprocity \cite{wassermanSNA}, meaning on the tourism network the
appetency to exchange tourists. The links in the network are
composed by 10\% bidirectional links and 30\% of asymmetric links.
If country \emph{j} has tourist arrivals from country \emph{i}, then the
probability that country \emph{i} has tourist arrivals from \emph{j} is only
$\frac{1}{4}$, so
the network is significantly directed. Notice also that 60\% of all
the pairs of countries are not connected to one another.

The tourism international network is a giant component, so that all
countries have a path or paths to any of the other countries. The
fact of being a giant network and having a small shortest path
length can imply fast transferring of knowledge and information.

On a directed network the nodes have \emph{in} and \emph{out}
degree, where the \emph{in} degree of a node $i$, $k_{in}(i)$, is
the number of nodes directed to node $i$, and the \emph{out} degree
of $i$, $k_{out}(i)$, is the number of nodes that $i$ is directed
to. The \emph{in} degree of a country is an indicator of its
attractiveness has a destination country, \emph{destination
attractiveness indicator}, which increases with the number of origin
countries that have flow of tourists to the destination on analyzes.
The \emph{out} degree of a country is an indicator of its emanation
has a tourism origin country, \emph{destination emanation
indicator}, which increases with the number of countries that the
country on analyzes has flow of tourists to.

A fundamental aspect of real-world networks is the degree
distribution \cite{barabasi1999}, representing the distribution of
the number of links of nodes. In binomial random graphs
\cite{mendesbook,Strogatz:ExploringComplexNetworks}, nodes have
similar degree, although many real-world networks have some nodes
that are significantly more connected than others, many of those are
scale free, having connectivity distributions that decay as a power
law. A probable mechanism for this occurrence is preferential
attachment \cite{barabasi1999}, meaning that nodes with high degree
are preferential. Network's topology displays the degree
distribution $P(k)$, probability that a node has degree $k$, which
applied to tourist arrivals - directed network
\cite{krapivsky-2001-86}- are studied two degree distribution
functions, $P_{in}(k)$ representing the probability that a node has
$k$ nodes directed to itself (probability of countries with tourism
from $k$ \emph{inbound} countries), $P_{out}(k)$ representing the
probability that a node has a total of $k$ edges to other nodes
(probability of countries with tourism to $k$ \emph{outbound}
countries). Most networks have a scale-free degree distributions
\cite{barabasi1999}, which have a power law tail $P(k) \sim
k^{-\theta}$.

\begin{figure}[!hbp]
  \includegraphics[width=8cm]{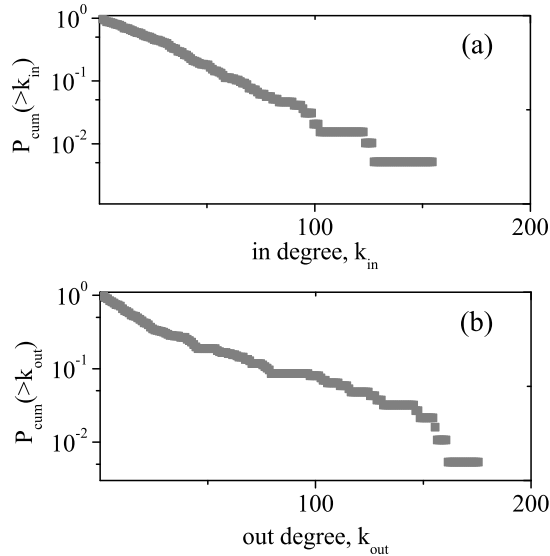}\\
  \caption{{\small Log-normal plot of degree distribution, for (a)
  \emph{in} degree $P(>k_{in})$ and (b) \emph{out} degree $P(>k_{out})$,
  with an exponential decay.}
  }\label{fig:dS}
\end{figure}

An exponential network is provided by the usual random graph, with
$P(k)$ decreasing exponentially fast, although scale-free networks
display a hub-like hierarchies, with $P(k)$ decreasing as a power
law \cite{albert-2002-74}. In our case, the \emph{in }and \emph{out}
degree distributions decrease exponentially fast, cumulative
distribution functions. On Fig. \ref{fig:dS} $(a)$ and $(b)$,
respectively $P_{in}(k)$ and $P_{out}(k)$. The topological network
does not displaying scale-free behavior, similar result on
\cite{Montis2005}.

The degree distributions decay is greatly faster than the power-law
degree distribution depicted in other social \cite{newman2001b},
technological \cite{barabasiInternet}, economic
\cite{chowell-2003-68}, and biological networks. Random networks are
described by growing model of random link assignment and models of
sublinear preferential attachment \cite{krapivsky-2000-85}.
Accordingly, the inbound and outbound degree distributions
reasonably suggest a growth model in which new connections are
chosen as a result of sublinear preferential attachment, where
higher degree countries are more likely to add new connections.
While, plausibly, destinations are chosen randomly, not being clear
if there is any advantage for making new connections with high
\emph{in} degree countries.

The weighed analysis is essential because of weights
heterogeneity. The network can be expressed by its adjacency matrix
$A=\{a_{ij}\}$, dimension N $\times$ N , where $a_{ij}=1$ if and
only if there is an edge from $i$ to $j$, and $a_{ij}=0$ otherwise.
The weighted adjacency matrix is $W=\{w_{ij}\}$, where $w_{ij}$
equals the flow from $i$ to $j$. Notice that $w_{ij}$ represents the
weight of the edge $i \to j$ and $w_{ji}$ represents the weight of
the edge $j \to i$, so $w_{ij}$ and $w_{ji}$ are different.

\begin{displaymath}
\xymatrix{
i \ar@<1ex>[dr]^{w_{ij}} \\
& j \ar@<1ex>[ul]^{w_{ji}} }
\end{displaymath}

The present network is asymmetric and weighted. The range of the
weights goes from $0$ to $19.369.677$ with an average value of
$81.813$, revealing a high heterogeneity of weights. See Fig.
\ref{fig:worldMap}

\begin{figure}
\begin{center}
\includegraphics[width=8cm]{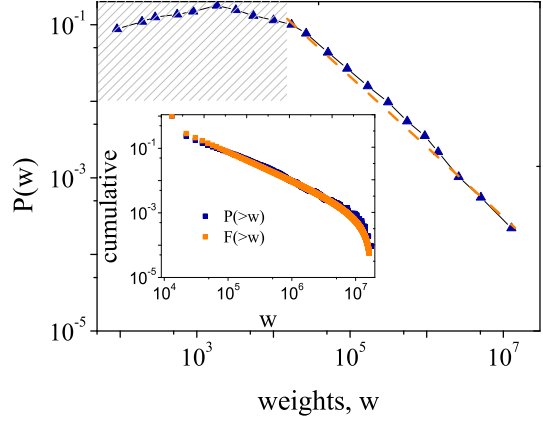}
\end{center}
\caption{{\small On the plot is showed the flow of tourists with a
scale-free behaviour $P(w) \sim w^{- \gamma}$ where $\gamma = 1.55$,
with dominance of hubs. The inner plot displays the cumulative of
$P(w)$ and the cumulative of the power-law with $\gamma = 1.55$, $F(w)$.}}
\label{fig:deS}
\end{figure}

The probability distribution function of the weights, $P(w) \sim
w^{- \gamma}$ has a power--law behavior, with exponent $\gamma = 1.55$,
see Fig. \ref{fig:deS}.

It is also relevant to study the strength of the nodes, which on a
directed network each node has \emph{in} strength, $s_{in}(i)$ (eq.
\ref{eq:strengthIN}), and \emph{out} strength, $s_{out}(i)$ (eq.
\ref{eq:strengthOUT}). It measures the strength of the nodes on
relation to the total weight of their connections. On the tourist
arrivals network
\emph{in} strength represents the \emph{inbound} tourism, and
\emph{out} strength represents the \emph{outbound} tourism. Strength
is a measure of centrality for weighted networks:

\begin{figure}
\begin{center}
\includegraphics[width=8cm]{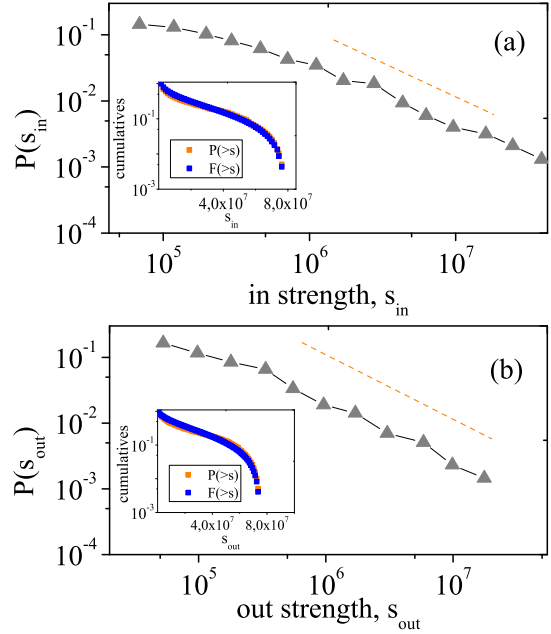}
\end{center}
\caption{{\small inbound distribution, $P(s_{in}) \sim
s_{in}^{\gamma_{in}}$ with $\gamma_{in}=0.9$, and outbound
distribution, $P(s_{out}) \sim s_{out}^{\gamma_{out}}$ with
$\gamma_{out}=0.95$, inner plots display their cumulative.}}
 \label{fig:strengthINOUTdegrees}
\end{figure}

\begin{equation} \label{eq:strengthIN}
s_{in}(i) = \sum_{j \in \upsilon (i)} w_{ij},
\end{equation}

\begin{equation} \label{eq:strengthOUT}
s_{out}(i) = \sum_{j \in \upsilon (i)} w_{ji}.
\end{equation}

The \emph{in} strength distribution and \emph{out} strength
distribution functions are also fitted by a power-law, respectively
$P(s_{in}) \sim s_{in}^{\gamma_{in}}$ and $P(s_{out}) \sim
s_{out}^{\gamma_{out}}$, where $\gamma_{out}=0.95$ and
$\gamma_{in}=0.9$, represented on Fig.
\ref{fig:strengthINOUTdegrees}.

Scale free networks have the ability to change scale in order to
meet any level of demand. Tourism, among economic sectors has one of
the fastest grow rates, and WTO forecasts that international
arrivals are expected to reach nearly $1.6$ billion \cite{yearbook}.
So, two consequences are expected, the network is growing due to a
scaling up, with an increase of flows intensity and/or due to a
scaling out by new connections between countries.

A power-law behavior of $P(w)$, $P(s_{in})$ and $P(s_{out})$ have a
strong structural meaning of the network, describing the way
weights, and strength centrality, \emph{inbound} and \emph{outbound}
tourism, are distributed. The weights and strengths range on a large
spectrum of values, and the heavy-tailed distribution implies that
nodes have a certain probability of having large strength values,
where the average of all intermediate values has no meaning.

\begin{figure}
\begin{center}
\includegraphics[width=8cm]{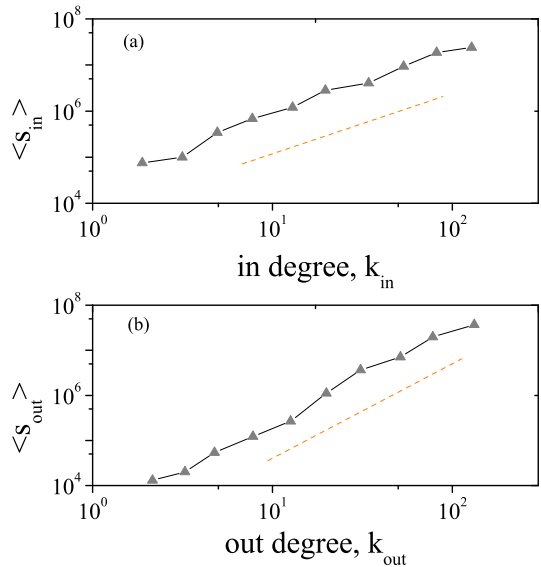}
\end{center}
\caption{{\small Intensity plays an important role on network
behaviour. The relation between degree and strength is closely
independent on (a) inbound tourism, $s(k_{in}) =
k_{in}^{\beta_{in}}$ with $\beta_{in}=1.1$ , but has a (b) strong
relation on outbound tourism, $s(k_{out}) = k_{out}^{\beta_{out}}$
with $\beta_{out}=1.75$. }}\label{fig:SD}
\end{figure}

The observations on topological and weighted network reveal
different structural results, therefore the relation of topological
and weighted flows is studied in more detail, $s(k_{in})$ and
$s(k_{out})$. The result is depicted on Fig. \ref{fig:SD}. On the
\emph{in} function:

\begin{equation}\label{eq:strengthindegree}
s(k_{in}) = (k_{in})^{\beta_{in}},
\end{equation}

where $\beta_{in}=1.1$. For $\beta = 1$ degree and weight are
independent \cite{barrat-2004-101}. So $S(k_{in})$ and $k_{in}$ are
close to independent, revealing a very small relation between them.
On the other side, $s(k_{out})$:

\begin{equation}\label{eq:strengthoutdegree}
s(k_{out}) = (k_{out})^{\beta_{out}},
\end{equation}

$\beta_{out}=1.75$, revealing a strong relation between \emph{out}
strength and \emph{out} degree. This means that \emph{outbound}
tourism increases with \emph{out} degree.

Interestingly, when analyzing the diversity of the market and its strength, comes out that
inbound and outbound tourism have distinguished outcomes on Fig. \ref{fig:SD}. Even so, both have a
power-law behaviour, $s(k) = k^{\beta}$, and unavoidable fluctuations. The diversification of outbound
markets ($>k_{out}$) has a strong and positive increase on total outbound tourism $s(k_{out}) = k_{out}^{\beta_{out}}$, with a
power of $\beta_{out}=1.75$, meaning that the flow grows $1.75$  faster than the degree.
On the relation between the inbound tourism and its market diversification,
$s(k_{in}) = k_{in}^{\beta_{in}}$ with $\beta_{in}=1.1$,
the relation is close to linear and it comes out that both quantities carry almost the same
information \cite{barrat-2004-101}.
It is concluded that the outbound tourism particularly benefits from market diversity.

\subsection{Degree-Degree Correlations}\label{sec:degreeCorrelations}

We turn now to question in which sense do countries couple with one another.
Is it in some sort of random choice, or is there a preference on the way they
link with each others, meaning a choice that makes some connections more
probable than others. In a social context is usually observed an assortative
mixing \cite{assortativity}, observed when the nearest neighbours of nodes with high degree
have also high degree. On economic, technological and biological context is
generally observed disassortative mixing, observed when the nearest neighbours
of nodes with high degree have low degree.

Degree-degree correlations, illustrate that
social networks have assortative mixing and technological and
biological networks behave more like disassortative. Worldwide
tourist arrivals  display
disassortative mixing, revealing the economic behaviour over social.
The weighted versus topological analysis of degree-degree
correlations shows that low (high) degree countries have their
inbound edges with large weight directed from countries with low
(high) degree.

In evolving network, degree-degree correlations are almost always
strong. To measure the correlation on the network over degree, one
may also study the average nearest-neighbors degree. This measures
the tendency of node $i$ to be connected to nodes with the same
degree,

\begin{equation} \label{eq:degreeCorrelation0}
k_{nn}' (i) = \frac{1}{k_{i}} \sum_{j \in \upsilon (i)}k_{j},
\end{equation}

where $\upsilon (i)$ denotes the set of neighbors of $i$.
Considering that our network is directed, we correlate the \emph{in}
degree of node $i$ with the \emph{out} degree of its neighbors,

\begin{equation} \label{eq:degreeCorrelation3}
k_{nn} (i) = \frac{1}{k^{in}_{i}} \sum_{j \in \upsilon
(i)}k^{out}_{j}.
\end{equation}

We can also average the over nodes of the same degree:

\begin{equation} \label{eq:degreeCorrelation1}
k_{nn} (k) = \frac{1}{NP(k)} \sum_{k_{i} = k}k_{nn}(i).
\end{equation}

This measure is also called \emph{associative mixing} if nodes with
high degrees have most of their neighbors with high degrees,
represented by a growth of $k_{nn} (k)$ with $k$. For a decreasing
of $k_{nn} (k)$  with $k$ it is denominated disassortative mixing.
This happens when nodes with high degrees have mainly neighbors with
low degree. The international tourism network displays
disassortative mixed. This behavior is mostly detected on
transportation networks, providing a pattern where the hubs connect
to the small degree nodes at the periphery of the network
\cite{assortativity}.

Degree-degree correlation for a weighted network is given by
\cite{Montis2005},

\begin{equation} \label{eq:degreeCorrelation4}
k^{w}_{nn} (i) = \frac{1}{s^{in}_{i}} \sum_{j \in \upsilon (i)}
w_{ji}k^{out}_{j}.
\end{equation}

\begin{figure}[!htb]
\begin{center}
\includegraphics[width=8cm]{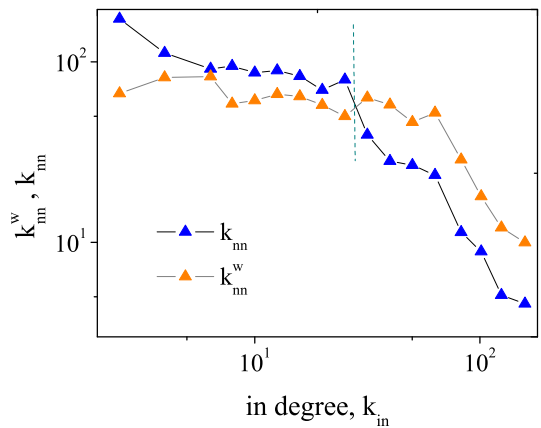}
\end{center}
\caption{{\small Log-log plot of \emph{in} degree -- \emph{out}
degree correlations, over \emph{in} degree, both unweighted
$k_{nn}(k)$ and weighted  $k_{nn}^w(k)$ correlations, over
$k_{in}$.}} \label{fig:degreeCorrelations}
\end{figure}

$k_{nn}^w (k)$ measures the local weighted average of neighbors
degree. The spectrum of the worldwide tourist flows on topological (equation
\ref{eq:degreeCorrelation1}) and weighted degree-degree correlations
(equation \ref{eq:degreeCorrelation4}) if represented on Fig.
\ref{fig:degreeCorrelations}.

\begin{figure}[!htb]
\begin{center}
\includegraphics[width=8cm]{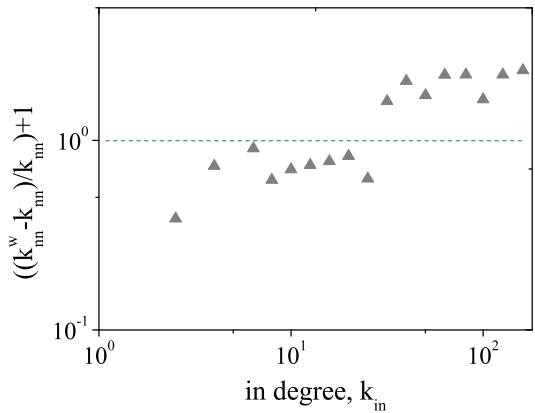}
\end{center}
\caption{{\small Comparing weighted and topological degree
correlation.
For low degrees
$k_{nn}^w(k)<k_{nn}(k)$ and for high degrees
$k_{nn}^w(k)>k_{nn}(k)$.
Low (high) degree
nodes have their edges with large weight directed from nodes with
low (high) degree. }}\label{fig:degreeCorrelationsCompare}
\end{figure}

For $k_{nn}^w(k)>k_{nn}(k)$ the edges with the larger weight are
directed to the neighbors with larger degrees, and
 $k_{nn}^w(k)<k_{nn}(k)$ the edges with the
larger weight are directed to the neighbors with lower degrees
\cite{Montis2005}. The weighted degree-degree correlation is
slightly decreasing (Fig. \ref{fig:degreeCorrelationsCompare}),
following the same behavior as the topological correlation, but with
a slower slop. For low degrees $k_{nn}^w(k)<k_{nn}(k)$ and for high
degrees $k_{nn}^w(k)>k_{nn}(k)$, meaning that low degree nodes have
their edges with large weight directed from nodes with low degree,
and high degree nodes have their edges with large weight directed
from nodes with high degree.

\section{Conclusion}\label{sec:conclusion}

This study provides a complementary perspective to the study of
international tourism. It is addressed the importance of weighted and directed
network measurements, by quantifying, on its topological and weighted structure,
the worldwide tourist arrivals network.

The research of tourist arrivals -- tourism, defined by the world
Tourism Organization, comprising the activities of persons traveling
to and staying in places outside their usual environment for not
more than one consecutive year for leisure, business and other
purposes1 - shows a scale-free behaviour on the weights covering $4$
orders of magnitude. It describes short travelling range to long
travels, on a global scale, surprisingly having affinity
correlations typical from technological and economic networks which
question the cultural backbone of tourism and travel. The scaling
behavior of tourism flows, on a power-law refers to the
self-organization of world trends, where disassortative correlations
particularly reveal the influence of economic flows and spread of
technologic and knowledge across international borders.

The scale-free nature of the weighted analyses contrary
to the random topology opens a new class of networks. This brings us to a more
general question; do highly heterogenic and directed real-world networks hide
some sort of preferential growing and hub-like structure on a random topological
structure?

\begin{acknowledgments}
J.I.L.M. acknowledges financial support from FCT
(SFRH/BD/19258/2004). J.F.F.M. acknowledges support from project
DYSONET-NEST/012911, and projects FCT:  FAT/46241/2002,
MAT/46176/2003. J.I.L.M. thanks Professor Carlos M. M. Costa and
Professor Lu\'is A. N. Amaral for research support and fruitful
discussions.
\end{acknowledgments}


\begin{thebibliography}{32}

\bibitem{milgram} J. Travers and S. Milgram, Sociometry 32 (1969) 425.

\bibitem{granovetter} M. S. Granovetter, The American Journal of
Sociology 78 (1973) 1360.

\bibitem{albert-2002-74} R. Albert, A.-L. Barabasi, Rev. Mod.
Phys. 74 (2002) 47.

\bibitem{goltsev-2002-51} S. N. Dorogovtsev, J. F. F. Mendes,
Adv. Phys. 51 (2002) 1079.

\bibitem{newman-2003-45} M. E. J. Newman, SIAM Review 45 (2003) 167.

\bibitem{jeong-2001-411} H. Jeong, S. P. Mason, A.-L. Barabasi, Z.N.
Oltvai, Nature 411 (2001) 41.

\bibitem{mendesbook} S. N. Dorogovtsev, J. F. F. Mendes, Evolution of networks: From
biological nets to the internet and WWW (Oxford Univ. Press, 2003).

\bibitem{Strogatz:ExploringComplexNetworks} S. H. Strogatz, Nature 410 (2001) 268.

\bibitem{wassermanSNA}
S. Wasserman, K. Faust, {\em Social Network Analysis} (Cambridge
University Press, 1994) ISBN 0521387078.

\bibitem{air} R. Guimer\`a, S. Mossa, A. Turtschi, L. A. N.
Amaral, Proc. Natl. Acad. Sci. USA \textbf{102}, 7794 (2004).

\bibitem{marketInvestment} G. Garlaschelli, S. Battiston, Physica
A 350 (2005) 491.

\bibitem{finantial} G. Caldarelli, S. Battiston, D. Garlaschelli, M.
Catanzaro, Lecture Notes in Physics 650 (2004) 399.

\bibitem{tibely-2006-370} G. Tibely, J.~-P. Onnela, J. Saramaki, K. Kaski, J.
Kertesz, Physica A 370 (2006) 145.

\bibitem{barabasiInternet} R. Albert, H. Jeong, A.-L Barabasi,
Nature 401 (1999) 130.

\bibitem{barabasi-2002-311} A.~L. Barabasi, H. Jeong, Z. Neda, E. Ravasz, A. Schubert, T.
Vicsek, Physica A 311 (2002) 3.

\bibitem{newman2001b} M. E. J. Newman, Proc. Natl. Acad. Sci. USA
98 (2001b) 404.

\bibitem{travel} D. Brockmann, L. Hufnagel, T. Geisel, Nature
439 (2006) 462.

\bibitem{Montis2005} A. {de Montis}, M. Barthelemy, A. Chessa, A.
Vespignani, Environment and Planning B: Planning and Design
34 (2007) 905924.

\bibitem{barrat-2004-101} A. Barrat, M. Barthelemy, A.
Vespignani, Proc. Natl. Acad. Sci. USA 101 (2004) 3747.

\bibitem{costa}
C. Costa, University of Surrey, England, PhD Thesis (1996).

\bibitem{tourismNetworks} L. Gibson, M. Hall, P. Lynch, R. Mitchell, A. Morrison, C.
Schreiber, \emph{Micro-Clusters and Networks: The Growth of
Tourism}, (Advances in Tourism Research Series, 2006).

\bibitem{tourismNetworksDriveTourism} H.-Y. Shih, Tourism Management
27 (2006) 1029.

\bibitem{rodolfo} R. Baggio, Physica A 379 (2007) 727.

\bibitem{rodolfo2} R. Baggio, Information Technology \& Tourism 8(1) (2006) 15.

\bibitem{yearbook} WTO, \emph{New Yearbook of Tourism Statistics} (World Tourism Organization Pbcn,
2006).

\bibitem{burt} R. S. Burt, \emph{Structural Holes: The Social Structure of
Competition} (Harvard University Press, 1995).

\bibitem{freeman} L. C. Freeman, Social Networks 1 (1979) 215.

\bibitem{goltsev-2003-67} A. V. Goltsev, S. N. Dorogovtsev, J. F. F.
Mendes, Phys. Rev. E 67 (2003) 026123.

\bibitem{park-2006-74} S. M. Park, B. J. Kim, Phys. Rev. E
74 (2006) 026114.

\bibitem{barrat-2004-92} A. Barrat, M. Barthelemy, A. Vespignani,
Phys. Rev. Lett. 92 (2004) 228701.

\bibitem{barabasi1999} R. Albert, A.-L. Barabási, Science
286 (1999) 509.

\bibitem{krapivsky-2001-86} P. L. Krapivsky, G. J. Rodgers, S.
Redner, Phys. Rev. Lett. 86 (2001) 5401.

\bibitem{chowell-2003-68} G. Chowell, J. M. Hyman, S. Eubank, C.
Castillo-Chavez, Phys. Rev. E 68 (2003) 066102.

\bibitem{krapivsky-2000-85} P. L. Krapivsky, S. Redner, F.
Leyvraz, Phys. Rev. Lett. 85 (2000) 4629.

\bibitem{assortativity} M. E. J. Newman, Phys. Rev. E 67 (2003) 026126.

\end{thebibliography}
\end{document}